\documentstyle[11pt,aaspp]{article}

\begin{document}

\begin{flushright} DPNU--94--12   \end{flushright}
 \vspace*{2cm}

\title{Fragmentation of a Magnetized Filamentary
Molecular Cloud Rotating around its Axis }
\author{Tomoaki M{\scriptsize ATSUMOTO}, Fumitaka N{\scriptsize AKAMURA},
and Tomoyuki H{\scriptsize ANAWA} \\
Department of Astrophysics, School of Science, Nagoya University,
Chikusa-ku, Nagoya, Aichi, 464-01 \\
E-mail(TM) matsu@a.phys.nagoya-u.ac.jp}

\centerline{(Received 1993 July 26; Accepted 1994 February 9)}

\centerline{\bf Abstract}

      The dynamical instability of a self-gravitating magnetized
filamentary cloud was investigated while taking account of rotation
around its axis.  The filamentary cloud of our model is supported
against self-gravity in part by both a magnetic field and rotation.
The density distribution in equilibrium was assumed to be a function of
the radial distance from the axis, $ \rho _0 (r) \, = \,
\rho _{\rm c} \, ( 1 \, + \, r ^2 / 8 H ^2 ) ^{-2} $, where
$ \rho _{\rm c} $ and $ H $ are model parameters specifying the
density on the axis and the length scale, respectively;
the magnetic filed was assumed to have both longitudinal ($z$-) and
azimuthal ($\varphi$-) components with a strength of
$ B _0 (r) \, \propto \, \sqrt{ \rho _0 (r) } $.
The rotation velocity was assumed to be $ v _{0\varphi} \, = \,
\Omega _{\rm c} \, r \, (1 \, + \, r ^2 / 8 H ^2 ) ^{-1/2} $.
We obtained the growth rate and eigenfunction numerically for (1)
axisymmetric $ ( m \, = \, 0 ) $
perturbations imposed on a rotating cloud with a longitudinal
magnetic field, (2) non-axisymmetric $ ( m \, = \, 1 ) $ perturbations
imposed on a rotating cloud with a longitudinal magnetic filed,
and (3) axisymmetric perturbations imposed on a rotating cloud
with a helical magnetic field.
The fastest growing perturbation
is an axisymmetric one for all of the model clouds studied.
Its wavelength is $ \lambda _{\rm max} \, \le \, 11.14 \, H $
for a non-rotating cloud without a magnetic field, and is shorter
when the magnetic filed is stronger and/or the rotation is faster.
For a rotating cloud without a magnetic filed
the most unstable axisymmetric mode is excited mainly by
self-gravity (the Jeans instability), while the unstable
non-axisymmetric mode is excited mainly by non-uniform rotation
(the Kelvin-Helmholtz instability).  The unstable non-axisymmetric
perturbation corotates with a fluid at $ r \, = \, 2 \, - \, 4 \, H $
and grows in time.  When the equilibrium magnetic field is helical,
the unstable perturbation grows in time and propagates along the
axis.  A rotating cloud with a helical magnetic field is less
unstable than that with a longitudinal magnetic field.


\bigskip
{\normalsize \hfill to appear in Publications of Astronomical
Society of Japan (vol.46, No.3)}
\bigskip

\clearpage

\noindent {\bf 1. \ \ Introduction}

     Recent observations have revealed that many molecular clouds
contain elongated filamentary structures.  These filamentary clouds
have dense cores as their internal constituents.
Most of the young stellar objects seem to be associated with dense cores.
The evolution, i.e., fragmentation, of a filamentary cloud, is therefore
interesting in relation to the early phase of star-formation processes.

     Filamentary molecular clouds are often associated with magnetic fields
which are perpendicular to the
cloud in the Taurus region (Moneti et al. 1984) and parallel to
the cloud in the Ophiuchus region (Vrba et al. 1976).
It has been suggested that the Orion A cloud has
helical magnetic fields around the cloud axis (Bally 1989;
Uchida et al. 1991).  In some
filamentary molecular clouds the velocity gradient has a component
perpendicular to the filament axis, which can be interpreted as
rotation around the axis (see, e.g., Olano et al. 1988;
Uchida et al. 1991; Tatematsu et al. 1993).
The energies of the magnetic
field and the rotation are likely to be comparable to the gravitational
energy.  These magnetic fields and rotation
may influence the fragmentation of the filamentary clouds.

    Since the early work of Chandrasekhar and Fermi (1953),
the linear stability of a cylindrical
gas cloud has been investigated extensively by Stodo\l kiewicz
(1963), Hansen et al. (1976), Nagasawa (1987),
Inutsuka and Miyama (1992), and Nakamura et al. (1993, referred to
as Paper I in the following).  These studies have taken account of
the effects of magnetic fields, the stiffness of the equation of state,
the rotation and
the collapse of the filamentary cloud in the radial direction.  However,
the stability of a rotating magnetized filamentary cloud against
fragmentation has not been
discussed.  Among the studies referred to above,
only Hansen et al. (1976)
took into account the rotation around the axis; however, they considered
neither the magnetic fields nor the fragmentation in the $ z $-direction.
We extend the model of Paper I to include rotation around
the axis and to discuss the fragmentation of a rotating filamentary cloud
with longitudinal or helical magnetic fields.

    The model and computation methods  are described in section 2.
In section 3 a rotating filamentary cloud with longitudinal magnetic
fields is shown to be unstable against axisymmetric and non-axisymmetric
perturbations.  In section 4 the stability of a rotating filamentary
cloud with helical magnetic fields is discussed.  In section 5 we
discuss the application of our stability analysis and compare our result to
the Kelvin-Helmholtz and Balbus-Hawley instabilities.

\bigskip

\noindent{\bf 2. \ \ Model and Computation Methods}

\medskip

\noindent{\it 2.1. \ \ Equilibrium Model}

     As a model of a filamentary molecular cloud,
we considered an infinitely long cylindrical gas cloud in equilibrium
in which the density and the magnetic field are uniform
along the $ z $ - axis in a cylindrical coordinate system
$ (r, \, \varphi , \, z ) $.  The model of Paper I
was extended so as to include rotation around the filamentary axis.
See, e.g., Bonnell et al. (1992) concerning rotation around an arbitary
axis.  They followed the fragmentation of a rotating finite-length cloud
using numerical simulations.

     The hydrostatic equilibrium of a filamentary molecular
cloud is described by
\begin{equation}
{dP _0 \over dr} \; + \; {d \over dr} \Bigl({B _{0\varphi} {}^2 \; + \;
B _{0z} {}^2 \over 8 \pi} \Bigr) \; + \; {B _{0\varphi} {}^2 \over 4 \pi r}
 + \; \rho _0 \Bigl( { v _{0\varphi} {} ^2 \over r } \, + \,
{d \psi _0 \over dr} \Bigr) \; = \; 0
\end{equation}
and
\begin{equation}
{1 \over r} {d \over dr} \Bigl(r {d \psi _0 \over dr} \Bigr)
 \;  = \; 4 \pi G \rho _0 \; .
\end{equation}
Here, $ P $, $ \rho $, $ v _\varphi $,
$ \psi $, $ G $, $ B _\varphi $, and $ B _z $
are the gas pressure, density, $ \varphi $-component
of the velocity, gravitational potential,
gravitational constant, and $ \varphi $- and
$ z $-components of the magnetic field, respectively.
Subscript 0 denotes the quantities in the unperturbed state.
The magnetic field of the $r$-component is assumed to vanish
in the unperturbed state,
$ \mbox{\boldmath $B$}_0 = (0, B _{0\varphi}, B _{0z}) $.
For simplicity, the filamentary gas cloud is assumed to be isothermal,
\begin{equation}
 P _0 / \rho _0 \; = \; c _{\rm s} {}^2 \; = \; {\rm const}.
\end{equation}
A solution satisfying equations (1) through (3) is expressed as
\begin{equation}
 \rho _0 \; = \; \rho _{\rm c}
\Bigl(1 \; + \; {r ^2 \over 8 H ^2} \Bigr) ^{-2} \; ,
\end{equation}
\begin{equation}
 \mbox{\boldmath $v$}_0 \; = \; (0, v _{0\varphi}, 0 ) \; = \;
\Bigl\{ 0 , r \Omega _{\rm c} \, ( 1 \, + \, r ^2 / 8 H ^2 ) ^{-1/2} , 0
\Bigr\} \; ,
\end{equation}
\begin{eqnarray}
 \mbox{\boldmath $B$}_0  & = & (0, B _{0\varphi}, B _{0z}) \; , \\
 B _{0\varphi}  &  = &  B _{\rm c} \sin \theta {r \over 2 \sqrt2 H}
 \Bigl(1 \; + \; {r ^2 \over 8 H ^2} \Bigr) ^{-3/2} \; , \\
 B _{0z} & = & B _{\rm c} \Bigl(1 \; + \; {r ^2 \over 8 H ^2} \Bigr)
 ^{-3/2} \sqrt{1 \; + \; \cos ^2 \theta {r ^2 \over 8 H ^2}} \; ,
\end{eqnarray}
and
\begin{equation}
 \psi _0 \; = \; 8 \pi G \rho _{\rm c} H ^2 \ln \Bigl( 1\; +
\; {r ^2 \over 8 H ^2} \Bigr) \; ,
\end{equation}
where
\begin{equation}
 4 \pi G \rho _{\rm c} H ^2 \; = \; c _{\rm s} {}^2 \; + \;
{B _{\rm c} {}^2 \over 16 \pi \rho _{\rm c}}(1 \; + \; \cos ^2 \theta) \;
+ \, 2 \Omega _{\rm c} {} ^2 H ^2  .
\end{equation}
All of the symbols with subscript c denote the values at $ r \; = \; 0 $.
The effective radius of this model is $ r \, = \, 2 \sqrt{2} H $.

     The density distribution of our model is the same as
those of Stod\'o\l kiewicz (1963), Ostriker (1964), Nagasawa (1987),
and Paper I.  Among these models,
our solution is the most general in the sense that our model incorporates
rotation around the axis as well as helical and longitudinal magnetic fields.
Our solution has five model parameters: $ \rho _{\rm c} , \; c _{\rm s},
 \; B _{\rm c} $, $ \theta $, and $ \Omega _{\rm c} $.
The $ \theta $ parameter denotes the ratio of the $\varphi$-
and $z$-components of the magnetic fields and is equal to the pitch angle
of the magnetic fields at $ r \; = \; \infty $,
$ \theta \; = \; \lim \limits_{r \rightarrow \infty} \tan ^{-1}
(B _{0\varphi} / B _{0z}) $.
When $ \Omega _{\rm c} \, = \, 0 $, our model reduces to that of Paper I.
When $ \theta \; = \; 0 $ and $ \Omega _{\rm c} \, = \, 0 $,
the magnetic field is parallel to the filament axis and
the model reduces to that of Stod\'o\l kiewicz's (1963) solution
for clouds with longitudinal magnetic fields.
When $ B _{\rm c} \; = \; 0 $, our solution reduces to
Ostriker's (1964) isothermal solution for non-magnetized clouds.
The sign of $ B _{\rm c} $ is taken to be positive in this paper
unless otherwise noted.
In this equilibrium model,
the ratios of the magnetic and centrifugal forces to the gas pressure
($ \alpha $ and $ \beta $) are spatially constant:
\begin{equation}
 \alpha \, \equiv \, {B _{0\varphi} {}^2 \; + \; B _{0z} {}^2
 \over 8 \pi P _0} \; = \; {B _{\rm c} {}^2 \over 8 \pi \rho _{\rm c} c _{\rm
s} {}^2}
\; = \; {\rm const}
\end{equation}
and
\begin{equation}
 \beta \, \equiv \, { 2 \Omega _{\rm c} {} ^2 H ^2 \over c _{\rm s} {} ^2 } \;
=
 \; {\rm const}.
\end{equation}
The condition for equilibrium [ equation (10)] thus reduces to
\begin{equation}
4 \pi G \rho _{\rm c} H ^2 \, = \,
c_{\rm s} {}^2 \left[ 1 \, + \, \alpha
\left( \frac{1 \, + \, \cos ^2 \theta }{2} \right)
\, + \, \beta \right] \; .
\end{equation}
In the following we take $H$ as the unit of length
and $ (2 \pi G \rho _{\rm c}) ^{-1/2} $ as the unit of time.
The model can thus be specified by three non-dimensional parameters:
$ \alpha $, $ \beta $, and $ \theta $.
In the subsequent sections we investigate the dependence
of the growth rate on $ \alpha $, $ \beta $, and $ \theta $.

\medskip

\noindent{\it 2.2. \ \ Perturbation Equations}

     We considered a perturbation of a small amplitude
superimposed on the equilibrium cloud
 described above. The linearized equation of motion
for the perturbation is given by
\begin{eqnarray}
 \rho _0 \Bigl\{ {\partial \mbox{\boldmath $v$}_1 \over \partial t} & + &
( \mbox{\boldmath $v$}_0 \cdot \nabla)
\mbox{\boldmath $v$}_1 \, + \, ( \mbox{\boldmath $v$}_1 \cdot \nabla )
\mbox{\boldmath $v$}_0 \Bigr\}
\; - \; {\mbox{\boldmath $j$}_0 \times
\mbox{\boldmath $B$}_1 \; + \; \mbox{\boldmath $j$}_1
\times \mbox{\boldmath $B$}_0 \over c} \nonumber \\
& \; & + \; \nabla P _1
 \; + \; \rho _1 \nabla \psi _0 \; + \; \rho _0 \nabla \psi _1
 \; = \; 0 \; ,
\end{eqnarray}
where $ \mbox{\boldmath $v$} $ and $ \mbox{\boldmath $j$} \;
= \; (c/4 \pi) \nabla \times \mbox{\boldmath $B$} $
are the velocity and electric current density, respectively; all
of the symbols with subscript 1 denote the changes in the quantities
due to the perturbation.
The equation of continuity, the Poisson equation, and the induction equation
are respectively expressed as follows:
\begin{equation}
 {\partial \rho _1 \over \partial t} \; + \; \nabla \cdot
(\rho _0 \mbox{\boldmath $v$} _1 \, + \,
\rho _1 \mbox{\boldmath $v$} _0 ) \; = \; 0 \; ,
\end{equation}
\begin{equation}
 \triangle \psi _1 \; = \; 4 \pi G \rho _1 \;
\end{equation}
and
\begin{equation}
 {\partial \mbox{\boldmath $B$}_1 \over \partial t} \;
 = \; \nabla \times (\mbox{\boldmath $v$} _1
 \times \mbox{\boldmath $B$}_0 \, + \, \mbox{\boldmath $v$} _0
 \times \mbox{\boldmath $B$} _1 ) \; .
\end{equation}
The perturbation is assumed to be isothermal,
\begin{equation}
 P _1 \; = \; c _{\rm s} {}^2 \rho _1 \; .
\end{equation}
We obtained the normal-mode solutions of equations (14) through (18),
in which all of the physical quantities change according to the form
\begin{equation}
 \rho _1 (r, \varphi, z, t) \; = \; \rho _1 (r) \exp  (-i \omega t \; + \;
i m \varphi \; + \; i k _z z ) \; .
\end{equation}
After some manipulation of equations (14) through (18), it can be rewritten as
\begin{equation}
 {d \over dr} \left(\matrix{ y _1 \cr y _2 \cr y _3
\cr y _4} \right) \; = \; \left(\matrix{A _{11} & A _{12} & A _{13}
& A _{14} \cr
A _{21} & A _{22} & A _{23} & 0 \cr 0 & 0 & 0 & A _{34}
\cr A _{41} & A _{42} & A _{43} & A _{44}} \right)
 \left(\matrix{y _1 \cr y _2 \cr y _3 \cr y _4} \right) \; ,
\end{equation}
where
\begin{equation}
 (y _1, \ y _2, \ y _3, \ y _4) \; = \; \Bigl(P _1 \ + \ {B _{1\varphi}
 B _{0\varphi} \ + \ B _{1z} B _{0z} \over 4 \pi}, \
 {i \rho _0 v _{1r} \over \omega},
 \ \rho _{\rm c} \psi _1, \ \rho _{\rm c} g _{1r} \Bigr)
 \; .
\end{equation}
The symbol $ g _{1r} (= d \psi _1/ dr) $
denotes the change in the gravitational acceleration
in the $ r $-direction.
The elements of matrix $ \mbox{\boldmath $A$} $
are given explicitly in appendix 1.

     The boundary conditions were set so that the perturbation is regular
at $ r \, = \, 0 $ and infinitesimal at $ r \, = \, \infty $.
Since the perturbation equation is singular at $ r \, = \, 0 $,
we obtained regular solutions near to $ r \, = \, 0 $ according to
the method of Paper I.
The asymptotic solutions in the region $ r \, \gg \, 4 H $ were
also obtained according to Paper I.

\medskip

\noindent {\it 2.3. \ \ Numerical Procedure}

     The eigenvalues and eigenfunctions were obtained essentially with the
methods of Paper I.  Namely, we integrated the perturbation equation
both from $ r \, = \, 0 $ and $ \infty $ for a given
$ \omega \, ( = \, \omega _r \, + \, i \omega _i ) $ and
checked whether the integrated solutions could be connected continuously
at a middle point.  When a solution satisfied the boundary conditions
at $ r \, = \, 0 $ and $ \infty $ for a given $ \omega $,
we regarded the given $ \omega $ as beeing an eigenvalue.

     We searched for eigenvalues using the bisection method.
When $ \Omega \, = \, 0$ or $ m \, = \, \theta \, = \, 0$,
the eigenvalue $ ( \omega ^2 ) $ is always real
and can be obtained by the usual bisection method.
When otherwise, we extended the bisection method for a complex eigenvalue.
The technical details concerning the extended bisection method are
given in appendix 2.

     Equation (20) has a singularity at $ v _{0 \varphi} \, = \,
r \omega _r / m $ when $ m \, \not= \, 0 $.  The result
depends on the integration path (see, e.g., Lin 1945a, b;
Kato 1987).   We integrated equation (20) over $ r $ along the real axis.
This integration path is correct only for unstable modes.
Although this integration path is not correct for damped modes,
we applied it for simplicity, since we are not much interested in
damped modes.

\bigskip

\noindent {\bf 3. \ \ Rotating Clouds with Longitudinal Magnetic Fields}

      In this section we consider the stability of a rotating cloud
with longitudinal magnetic fields.
The parameter $ \theta $ is fixed to be $ 0 ^\circ $ in this section,
except where otherwise stated.

\medskip

\noindent{\it 3.1. \ \ Sausage $ ( m \, = \, 0 ) $ Mode}

      In the beginning we consider the sausage mode $ ( m \, = \, 0 ) $
instability of a rotating  non-magnetized
filamentary cloud.
The eigenvalue $ ( \omega ) $ is purely imaginary for the unstable mode.
 Figure 1 shows the dispersion relation for the
model of $ ( \alpha , \, \beta ) \, = \, ( 0 , \, 1) $
by the thick curve.
The abscissa and ordinate are the non-dimensional wavenumber $ (k _z H) $
and the non-dimensional growth rate
$ ( \omega _i / \sqrt{ 2 \pi G \rho _{\rm c} } ) $, respectively.
There is only one unstable sausage mode for the models with
$ \alpha \, = \, 0 $, although there are some unstable sausage modes
for $ \alpha \, > \, 0 $ (Paper I).  The sausage mode is unstable only when
the wavenumber is smaller than a critical one, $ k _{z, \, {\rm cr}} H \, = \,
0.960 $. The growth rate has its maximum,
$ \omega _{i, \, {\rm max}} \, = \, 0.606 \, \sqrt{ 2 \pi G \rho _{\rm c} } $,
at $ k _z H \, = \, 0.467 $.
The dashed and thin curves denote the
growth rate of the most unstable mode for $ ( \alpha , \, \beta ) \,
= \, ( 0 , \, 0 ) $ and $ ( \alpha , \, \beta ) \, = \,
( 1 , \, 0 ) $, respectively.  The rotation as well as
the longitudinal magnetic fields increase the maximum growth rate
$ ( \equiv \, \omega _{i, \, {\rm max}} ) $,
the wavenumber of the fastest growing perturbation
$ ( \equiv \, k _{z, \, {\rm max}} ) $, and the critical wavenumber
$ ( \equiv \, k _{z, \, {\rm cr}} ) $.
This is because both rotation around the axis and longitudinal magnetic
fields support the gas against gravity in the radial direction,
but does not operate in the $z$-direction.
The gas temperature is lower for fixed $ \rho _{\rm c} $ and $ H $
when $ \alpha $ and $ \beta $ are larger
$ \lbrack $ see equation (10) $ \rbrack $.  Accordingly the Jeans length,
the typical length scale for fragmentation due to self-gravity,
is shorter.
The wavenumber of the fastest growing perturbation is almost the same
for $ ( \alpha , \, \beta ) $ = $ ( 0, \, 1 ) $
and $ ( 1 , \, 0 ) $.  This implies that $ k _{z, \, {\rm max}} $ is
a function of $ c _{\rm s} $ for $ \theta \, = \, 0 ^\circ $.  The maximum
growth rate is smaller for $ ( \alpha , \, \beta ) \,
= \, ( 0 , \, 1 ) $ than for
$ ( \alpha , \, \beta ) \,
= \, ( 1 , \, 0 ) $.  This means that the Parker instability
increases the growth rate when a magnetic field is present.

       Figure 2 shows the dependence of the growth rate on $ \beta $
for $ \alpha \, = \, 0 $.  As $ \beta $ increases, $ \omega _{i, \, {\rm max}}
$,
$ k _{z, \, {\rm max}} $, and $ k _{z, \, {\rm cr}} $ increase.  The wavenumber
of the
the fastest growing perturbation can be approximated by
\begin{equation}
k _{z, \, {\rm max}} H \, = \, 0.72 \, [ (1 \, + \, \alpha \, + \, \beta )
^{1/3} \, - \, 0.6 ]
\end{equation}
for $ \theta \, = \, 0 ^\circ $.  Figure 3 shows the accuracy
of equation (22).  The upper panel shows $ k _{z, \, {\rm max}} $ as a function
of $ \alpha \, + \, \beta $.  Equation (22) is drawn by the curve and
the numerically obtained values $ ( k _{z, \, {\rm max}} ) $
are plotted with filled circles.  The lower panel
shows the deviation from equation (22).  Equation (22) gives a good
estimate for $ k _{z, \, {\rm max}} $ with an error of 2\% for
$ 0 \, \le \, \alpha \, + \, \beta \, \le \, 6 $.
Equation (22) is used for a comparison with a filamentary molecular
cloud, Orion A, by Hanawa et al. (1993).  See Hanawa et al. (1993)
for an application of equation (22).
The approximate dispersion relation is given by
\begin{eqnarray}
\omega ^2 \,  & = & \, - 4 \pi G \rho _{\rm c} \, {k _z H \over 1 \, + \, k _z
H }
                    \, { 0.89 \, + \, 1.4 \alpha \over 1 \, + \, 1.25 \alpha }
                    \, + \, c _{\rm s} {}^2 k _z {}^2 \nonumber \\
              & = & \, - 4 \pi G \rho _{\rm c} \,
                    \left( \frac{ k_z H }{ 1 \, + \, k_z H }
                           \frac{ 0.89 \, + \, 1.4 \alpha }
                                { 1 \, + \, 1.25 \alpha }
                        \, - \, \frac{ H ^2 k_z{}^2 }
             { 1 \, + \, \alpha \, + \, \beta }
                    \right)
\end{eqnarray}
for the most unstable sausage mode for all the models of
$ \theta \, = \, 0 ^\circ $.
The upper and lower expressions of equation (23) are
equivalent [see equation (13)].
The second term on the right-hand side of equation (23)
denotes the stabilization due to thermal pressure,
i.e., the dispersion relation of the sound wave.
When $ \alpha $ and $ \beta $ are larger, $ c_{\rm s} $ is lower,
and, accordingly, the growth rate is larger.
The maximum growth rate obtained from
equation (23) agrees with the numerical results with an error of 5\%
for $ 0.5 \, \le \, \alpha \, \le \, 10 $ and $ 0 \, \le \, \beta \,
\le \, 10 $ , and for
$ \alpha \, = \, 0 $ and $ 0 \, \le \, \beta \, \le \, 2$.

      Figure 4 shows the fastest growing perturbation for
$ ( \alpha , \, \beta ) \, = \, ( 0 , \, 1 ) $.
The density distribution in the $ r - z $ plane is indicated by the grey
scale and the velocity field is denoted by arrows.
The amplitude of the perturbation is taken to be
$ \rho _1 / \rho _0 \, = \, \varepsilon \, \cos \, ( k _z z ) $
at $ r \, = \, 0 $.  To emphasize the density contrast we have taken
$ \varepsilon \, = \, 0.65 $ in figure 4.  The relative density perturbation
$ ( \vert \rho _1 / \rho _0 \vert ) $ has its maximum on the axis
$ ( r \, = \, 0 ) $.   The velocity perturbation is dominant in
the $ z $-direction.  The velocity perturbation in the $ z $-direction
has its maximum, $ v _{1z} \, = \, -0.941 \, \varepsilon \, c _{\rm s} \,
\sin \, ( k _z z ) $, on the axis and monotonically decreases as $ r $
increases.  The velocity perturbation in the $ r $-direction
has its maximum, $ v _{1r} \, = \, -0.128 \, \varepsilon \, c _{\rm s} \,
\cos \, ( k _z z ) $, at $ r \, = \, 2.73 \, H $.
The $ \varphi $-component of the velocity is shown in the right panel.
The solid and dashed curves denote the velocity in the perturbed state
and in equilibrium, respectively.
The change in the
rotation velocity is small with a maximum,
$ v _{1\varphi} \, = \, 0.193 \, \varepsilon
 \, c _{\rm s} \, \cos \, ( k _z z ) $,  at $ r \, = \, 1.76 \, H $.

      The eigenfunction shown in figure 4 is similar to that of the
the fastest growing perturbation
for a magnetized non-rotating cloud (see figure 8
of Paper I).  A filamentary cloud supported in part by either rotation
or longitudinal magnetic fields fragments mainly in the $ z $-direction.
We expected in Paper I that a disk perpendicular to the rotation axis
is formed by the nonlinear growth of the perturbation.
Tomisaka (1993) and Nakamura et al. (1993, private communication)
showed this by numerical simulations for magnetized filamentary
clouds.  We also followed
the nonlinear evolution of a rotating filamentary cloud
 with numerical simulations,
and confirmed the formation of a small disk perpendicular to the
axis.  The results of these numerical simulations will be reported
in a future paper.

      Figure 5 is the same as figure 4, except for $ ( \alpha , \, \beta )
\, = \, ( 0.5 , \, 0.5 ) $.
The velocity perturbation is dominant in
the $ z $-direction.  The velocity perturbation in the $ z $-direction
has its maximum, $ v _{1z} \, = \, -0.886 \, \varepsilon \, c _{\rm s} \,
\sin \, ( k _z z ) $, on the axis when $ \rho _1 / \rho _0 \, = \,
\varepsilon \, \cos \, ( k _z z ) $.
The velocity perturbation in the $ r $-direction
has its maximum, $ v _{1r} \, = \, -0.175 \, \varepsilon \, c _{\rm s} \,
\cos \, ( k _z z ) $, at $ r \, = \, 2.87 \, H $.
The change in
the rotation velocity is at most $ v _{1\varphi} \, = \, 0.101 \, \varepsilon
 \, c _{\rm s} \, \cos \, ( k _z z ) $ at $ r \, = \, 1.63 \, H $.
The characteristics of the most unstable sausage
mode for $ ( \alpha , \, \beta ) \, = \, ( 0.5 , \, 0.5 ) $
are between those for
$ ( \alpha , \, \beta ) \, = \,
( 1 , \, 0 ) $ and $ ( 0 , \, 1 ) $,
except for the generation of $ B _{\varphi} $.
When $ \alpha \beta \, \not= \, 0 $, the magnetic field is twisted
due to a change in the rotation velocity
$ \lbrack  B _{1\varphi} \, = \, -0.249 \, \varepsilon \, \sin \,
( k _z z ) $ at $ r \, = \, 1.51 \, H $, at most while
$ B _{0z} \, = \, 3.545 \, (1 \, + \, r ^2 / 8 H ^2 ) ^{-1} \rbrack $.

\medskip

\noindent {\it 3.2. \ \ Kink $ ( m \, = \, 1 ) $ Mode}

      Rotating filamentary clouds are also unstable against non-axisymmetric
perturbations.
We found five modes, three of which are pure oscillations
$ ( \omega _i \, = \, 0 ) $ for
any $ k _z $.  The remaining two modes have complex conjugate
eigenfrequencies.
Figure 6 shows the dispersion relation of the unstable
kink mode
$ ( m \, = \, 1 ) $ for $ \alpha \, = \, 0 $ and $ \beta \, = \, 1 $.
The real and imaginary
parts of the eigenfrequency are denoted by the dashed and solid curves,
respectively.
The dispersion relation
is symmetric with respect to $ k _z $, i.e., $ \omega ( - k _z ) \, =
\, \omega ( k _z ) $. We thus restrict ourselves to $ k _z \, > \, 0 $
in the following.
The three horizontal dotted lines
in the figure denote the angular velocities , $ \Omega $,
at $ r \, = \, 2, \, 2 \sqrt{2} $, and 4,respectively.
The unstable modes have a corotation point where the phase velocity coincides
with the fluid velocity $ \lbrack \omega _r / m \, = \, \Omega (r) \rbrack $.
The corotation point is located near to the effective radius,
$ r \, = \, 2 \sqrt{2} H $, when the growth rate is large.
The growth rate is very small
$ ( \omega _i \, < \, 1 \times 10 ^{-4} \, \sqrt{2 \pi G \rho _{\rm c}}) $
in the region $ \vert H  k _z \vert \, < \, 0.04 $, and large in the
region $ 0.1 \, \la \, H k _z \, < 0.477 $.
Because of the analytical
and numerical difficulty at the corotation point
$ \lbrack \omega _r \, = \, m \Omega ( r ) \rbrack $,
we could not find whether $ \omega _i \, = \, 0 $
in the region $ \vert H k _z \vert \, \le \, 0.04 $.
When $ k _z $ is larger than the critical value $ ( k _{z, \, {\rm cr}} \, =
\, 0.477 \, H ^{-1} ) $, the growth rate is again small
$ ( \omega _i \, < \, 10 ^{-2} \, \sqrt{2 \pi G \rho _{\rm c} } ) $.

     The mode having $ m \, = \, - 1 $ is physically the same as that having
$ m \, = \, 1 $, except for the sign of $ \omega _r $.
When $ m \, = \, -1 $, a mode with $ \omega _r \, < \, 0 $
becomes unstable and a mode with $ \omega _r \, > \, 0 $ is a pure
oscillation.  Only when the fluid velocity $ ( = \, \Omega ) $
and the phase velocity of the wave $ ( = \, \omega _r /m) $ have
the same sign can the mode be unstable.

      Figures 7 and 8 are the same as figure 6, except for
$ ( \, \alpha , \, \beta ) \, = \, ( \, 0, \, 0.5 ) $ and
$ ( \, 0 , \, 2 ) $, respectively.
The growth rate is larger when $ \beta $ is larger.  The non-rotating
filamentary cloud without magnetic fields $ ( \alpha \, = \, \beta \,
= \, 0 )$ has two neutrally stable kink modes with
 positive and negative real frequencies (see also Nagasawa 1987).
Only the mode having a corotation point becomes unstable; the other mode
remains pure oscillatory.
We can thus conclude that this instability is due to the resonance of the wave
to
the fluid rotation, and, accordingly, is of the Kelvin-Helmholtz type.
Note that non-rotating magnetized filamentary clouds are
unstable against the kink mode because of the Parker instability
(Paper I).

       When $ \vert k _z \vert $ is small,
the growth rate is very small (even if it is positive) and
the oscillation frequency is almost independent of $ \beta $.
For long-wavelength perturbations $ v _z $ has a large amplitude
in the region very far from the axis $ ( r \, \ga \, 10 \, H ) $.
For $ \vert k _z \, \vert \, \ge \, k _{z, \, {\rm cr}} $ the growth rate of
the
kink mode is very small $ ( \omega _i \, < \, 1 \times 10 ^{-2} ) $.
The critical wavenumber is $ H k _{z, \, {\rm cr}} \, = \, 0.301 $,
0.477, and 0.788
for $ ( \alpha , \, \beta ) $ = $ ( 0, \, 0.5 ) $,
$ (0, \, 1) $, and $ ( 0 , \, 2 ) $, respectively.
The growth rate has its maximum
[$ \omega _i / \sqrt{ 2 \pi G \rho _{\rm c} }
 \, = \, 0.0564$, 0.135, and 0.255] at $ H k _z \, =  \, 0.225$,
0.323, and 0.478, for $ ( \alpha , \, \beta ) $ = $ ( 0, \, 0.5 ) $,
$ (0, \, 1) $, and $ ( 0 , \, 2 ) $, respectively.

     Figure 9 is the same as figure 4, except for the most unstable
kink perturbation for $ ( \alpha , \, \beta ) \, = \, ( 0 , \, 1 ) $.
Figures 9(a) and (b) denote the cross section in the $ r \, - \, z $
plane and that of $ z \, = \, 0 $, respectively.  The growth rate
and wave number are $ \omega _i \, = \, 0.135 \, \sqrt{2 \pi G \rho _{\rm c} }
$
and $ H k _z \, = \, 0.323 $, respectively.  The amplitude of the
perturbation is normalized so that the density perturbation
is at most $ \rho _1 / \rho _0 \, = \, 0.5 $.  When the kink mode
is excited, the density ridge is twisted, as can be seen in
figure 9.

      Figure 10 is the same as figure 6, except for
$ ( \alpha , \, \beta ) $ = $ ( 0.5 , \, 0.5 )  $.
This model is intermediate between
those of $ (\alpha , \, \beta , \, \theta ) $
= $ ( 1 , \, 0, \, 0 ^\circ ) $ and $ ( 0 , \, 1 , \, 0 ^\circ ) $,
the former of which is excited by the Parker instability.
For a fixed value of $ \alpha \, + \, \beta $ the kink mode has a larger
growth rate when $ \alpha $ is larger.

\bigskip

\noindent {\bf 4. \ \ Rotating Cloud with Helical Magnetic Fields}

      In this section we discuss the sausage mode instability of
rotating filamentary clouds with helical magnetic fields
$ ( \alpha \beta \theta \, \neq \, 0 ) $.

      Figure 11 shows the dispersion relation of the most
unstable sausage mode for $ \theta \, = \, 30 ^\circ , \,
60 ^\circ $, and $ \, 90 ^\circ $,
while $ \alpha \, = \, \beta \, = \, 0.5 $.
The real (figure 11a) and imaginary (figure 11b) parts of the eigenfrequency
are shown as a function of $ k _z $.  The sausage mode
grows and propagates with a phase speed of $ \omega _r / k _z $
in the $ z $-direction.
The eigenfrequency of the damped mode is the complex conjugate of
that of the unstable mode.  The eigenfrequency is antisymmetric
with respect to $ k _z $, i.e.,
$ \omega ( - k _z ) \, = \, - \omega ( k _z ) $.
The phase velocity is thus always positive for both the unstable
and damped modes.
When $ \theta \, < \, 0 $, the real part of the eigenfrequency is
negative, $ \omega _r \, < \, 0 $.  Note that $ \Omega _{\rm c} $
is taken to be positive in our computations.
The phase velocity is $ \omega _r / k _z \, \simeq \, 0.025 \,
H \sqrt{2 \pi G \rho _{\rm c} } $
for $ \theta \, = \, 30 ^\circ $.
The phase velocity is faster for a larger $ \theta $.

      As $ \theta $ increases, $ \omega _{i,{\rm max}} $, $ k _{z, \, {\rm
max}} $,
and $ k _{z, \, {\rm cr}} $ decrease, and, accordingly, the cloud becomes less
unstable.
This is due to an increases in the sound speed, $ c _{\rm s} $,
for fixed $ \alpha $ and $ \beta $ [ see equation (13)].
Both when the magnetic field is longitudinal and helical,
the most unstable mode is mainly excited by
a self-gravitational instability.
The eigenfrequency $ \omega $ is complex for $ \beta \theta \; \not= \; 0 $,
while it is pure imaginary or real for $ \beta \theta \; = \; 0 $.
The perturbation therefore grows in time and propagates in the $ z $-direction
only when the cloud rotates and is threaded by a helical magnetic field.

        Figure 12 shows the fastest growing perturbation for
$ ( \alpha , \, \beta , \, \theta ) \, = \, (0.5, \, 0.5 , 60 ^\circ ) $.
Figure 12 is similar to figure 5, except for the phase shift in the
$ z $-direction.  The density perturbation is proportional to
$ \rho _1 \, \propto \,
\cos ( k _z z \, - \, \omega _r t \, + \delta _\rho ) $,
where the phase shift, $ \delta _\rho $, is a function of $ r $,
and is taken to be $ \delta _\rho \, = \, 0 $ at $ r \, = \, 0 $.
The phase shift is large for the azimuthal components
of the velocity and magnetic fields, e.g.,
$ v _{1\varphi} \, = \, 0.141 \, \varepsilon \cos \, ( k _z z \, - \,
\omega _r t \, + \, 0.561 ) $ at $ r \, = \, 1.73 \, H $ and
$ B _{1\varphi} \, = \, 0.682 \varepsilon \sin ( k _z z \, - \,
\omega _r t \, + \, 0.356 ) $ at $ r \, = 1.25 \, H $.
The upper panel of figure 12 shows the $ z$-dependence of $ v _{1\varphi} $
(the dashed curve) at $ r \, = \, 1.73 \, H $
and $ B _{1\varphi} $ (the solid curve) at $ r \, = \, 1.25 \, H $,
both of which have their maximum amplitudes there.

   The phase shifts are related to the propagation of the wave.
Consider a velocity perturbation proportional to $ v _{1r} \,
\propto \, \cos \, ( k _z z ) $ and
$ v _{1z} \, \propto \, \sin \, ( k _z z) $.  The velocity perturbation
changes $ v _\varphi $ and $ B _\varphi $ according to the
azimuthal components of the equation of motion and the induction equation.
When $ \theta \, = \, 0 $, the changes in the azimuthal components
have the dependence $ B _{1\varphi} \, \propto \, \sin \, (k _z z ) $
and  $ v _{1\varphi} \, \propto \, \cos \, ( k _z z ) $ (see section 3).
When $ \beta \, = \, 0 $, they have the dependence
$ B _{1\varphi} \, \propto \, \cos \, (k _z z) $ and
$ v _{1\varphi} \, \propto \, \sin \, (k _z z) $ (see Paper I).
When $ \beta \theta \, \not= \, 0 $, both types of the above-mentioned
perturbations are produced.  These changes in $ v _\varphi $
and $ B _\varphi $ modify $ v _r $ and $ v _z $ according to the equation of
motion.  As a result, velocity perturbations
given by $ v _{1r} \, \propto \, \sin \, ( k _z z ) $ and $ v _{1z} \,
\propto \, \sin \, ( k _z z ) $ are also induced
and the velocity perturbations apparently propagate.

\bigskip

\noindent{\bf 5. \ \ Discussion}

      As shown in the previous sections, a rotating magnetized filamentary
cloud suffers from various instabilities.  This is in part because
our model includes rotation as well as magnetic fields
as well as the self-gravity of the gas.
In this section we discuss the relationship between our model
and other theoretical studies conserning the instability of a rotating gas
cloud.
In the last paragraph we also discuss the application of our model
to TMC 1 (Taurus Molecular Cloud 1).

\medskip

\noindent{\it 5.1. \ \ Relationship to the Kelvin-Helmholtz Instability
and the Balbus-Hawley Mechanism}

      Since the angular velocity is non-uniform in our model, we suppose
that the Kelvin-Helmholtz instability may be involved in the stability
of our model cloud.  As discussed by Glatzel (1987a,b),
 the resonant interaction of
two neutral modes produces a couple of growing and damped modes and
causes an angular momentum transfer across the corotation point.
These characteristics are shared with the unstable kink mode for a
rotating cloud without a magnetic field.  As shown in subsection 3.2,
the growing and damped modes appear in pairs.  We confirmed that
the angular momentum is transferred from the region inside the corotation
point to that outside the corotation point when the unstable kink
mode is excited.  In the growing mode, $ \rho _1 $ and $ v _{1\varphi} $
are anti-correlated $ ( \rho _1 {}^* \cdot v _{1 \varphi} \, < \, 0 ) $
inside the corotation point and are positively correlated
$ ( \rho _1 {}^* \cdot v _{1\varphi} ) $ outside the corotation point,
where $ \rho _1 {} ^* $ denotes the complex conjugate of $ \rho _1 $.
This means that the central region loses angular momentum and the
outer part receives it.  The angular-momentum transfer produces an energy
excess, which causes the perturbation to grow.

      The energy excess due to the angular momentum transfer is
proportional to the difference in the angular velocity across the
corotation point.  Correspondingly, the growth rate of the Kelvin-Helmholtz
instability is bounded by the angular velocity gradient,
$ \omega _i \, \le \, {\rm max} \, \vert (1/2) r d \Omega / d r \vert $
(Sung 1974; Hanawa 1986, 1987; Fujimoto 1987), when the self-gravity of
the gas and magnetic fields are negligibly small.  In our model,
the angular velocity gradient,
\begin{equation}
r { d \Omega \over d r } \; = \; - \, \Omega _{\rm c} \,
{ r ^2 \over 8 H ^2 } \, \Bigl( 1 \, + \, { r ^2 \over 8 H ^2 } \Bigr)
^{-3/2} \; ,
\end{equation}
has its maximum, $ 0.385 \, \Omega _{\rm c} $ at $ r \, = \, 4 H $,
and a somewhat smaller value $ 0.355 \, \Omega _{\rm c} $ at
the effective radius, $ r \, = \, 2 \sqrt{2} \, H $.
It may be not a chance coincidence that the growth rate of the kink mode
is large when the angular velocity gradient is large at the corotation
point.

      The unstable kink mode might be excited in part by another effect
related to a large $ \beta $.
When $ \beta $ is large, the sound speed $(c _{\rm s})$
and thermal pressure are low [ see equation (13)].
The low gas pressure strengthens the effect of the self-gravity relatively and
suppresses the Jeans instability less effectively.  This effect is
appreciable in the axisymmetric mode when $ \beta \, \ga \, 1 $.
To investigate
this effect on the kink mode, we made an experimental model
in which the right-hand side of equation (16)
was multiplied by a factor of $ 0 \, \le \, \varepsilon \, \le 1 $,
$  \triangle \psi _1 \; = \; 4 \pi G \rho _1 \; $.
By decreasing $ \varepsilon $ from unity, the change in the gravitational
potential was reduced artificially.  As $ \varepsilon $ decreases,
$ \omega _r $ increases and $ \omega _i $ decreases.
The increase in $ \omega _r $ is ascribed to a reduction in the self-gravity,
since the self-gravity lowers the propagation of the sound wave.
The increase in $ \omega _r $ shifts the corotation point inwards,
and, accordingly, decreases the angular velocity gradient thereof.
The decrease in $ \omega _i $ may be in part due to a decrease
in the angular velocity gradient at the corotation point.  Thus,
unfortunately, we could not decouple the effect of low thermal pressure
on the Jeans instability from other effects of large $ \beta $.

     When $ \alpha \beta \, \not= \, 0 $, the unstable sausage mode
is in part due to the Balbus-Hawley~(1991) mechanism, although it is mainly
due to the Jeans instability.
Balbus and Hawley showed that a differentially rotating disk is highly unstable
if it is threaded by a vertical magnetic filed.
In such a disk, an outwardly displaced fluid element tries to enforce
a rigid rotation, and thus rotates too fast for its new radial location.
Then, the excess centrifugal force drives the element still faster outward.
As can be seen in subsection 3.1, the
unstable sausage mode generates $ B _r $ and $ B _\varphi $.  The
twisted magnetic fields transfer angular momentum along the field
line.  We confirmed that the fluid element displaced inward loses
angular momentum due to magnetic drag
when the unstable sausage mode is excited for
$ \alpha \beta \, \not= \, 0 $.  Because of the angular momentum
gain, the fluid element contracts further
due to a decrease in the centrifugal force.
In order to evaluate the contribution of the Balbus-Hawley mechanism to
the unstable sausage mode,
we again computed the artificial models in which the change in the
gravitational
potential was attenuated by a factor of $\varepsilon$.
The growth rate, $\omega_i$, decreases along with a decrease
in $\varepsilon$, and
diminishes at $\varepsilon\,=\,0$.
We thus conclude that the unstable sausage mode is excited mainly by
self-gravity.
When $\varepsilon\,=\,0$, the increase in the magnetic pressure near to the
axis
$(r\,=\,0)$ stabilizes our model cloud against the Balbus-Hawley mechanism
for $m\,=\,0$.
We confirmed that the sausage mode is stabilized against the Parker instability
by the same mechanism for $(\alpha,\,\beta)\,=\,(1,\,0)$
when $\varepsilon\,=\,0$.

\medskip

\noindent{\it 5.2. \ \ Comparison with Habe et al.'s Simulation}

     Here, we comment on the collapse of a self-gravitating
cloud triggered by torsional Alfv\'en Waves.  Habe et al. (1991) showed
by a numerical simulation that a cloud with helical magnetic
fields collapse more easily than one with longitudinal magnetic
fields.
In their simulation,
helical magnetic fields are generated by rotation and cause
a rotating cloud to collapse due to magnetic pinching.
It seems as if our stability analysis contradicts their simulation,
since  a rotating cloud with helical magnetic fields is less unstable
than that with longitudinal magnetic fields in our analysis.
The apparent contradiction comes from a difference in the initial
models.  While our initial model is in equilibrium, Habe et al's (1991)
started their numerical simulation from a cloud threaded by helical
magnetic fields.  Their initial model is not in equilibrium.
And our stability analysis cannot be applied directly to their initial model.

\medskip

\noindent{\it 5.3. \ \ Application to TMC 1}

    Finally, we discuss the application of our model to
a filamentary molecular cloud.
TMC 1 contains 5 dense cores which are
more or less regularly spaced with an average separation of
$ 4 \arcmin $.  Since the apparent diameter is $ 2 \arcmin $,
the ratio of the average separation to the filament diameter is
estimated to be $ \approx 2 $.  Substituting this value into
our model, we obtain $ H k _{\rm max} \, \simeq \, 0.63 $.  This means
that the magnetic and/or centrifugal forces are comparable to the
gravitational force $ ( \alpha \, + \, \beta \, \approx \, 1 ) $.
This is consistent
with the velocity structure of TMC 1.  Olano et al. (1988) found a
velocity gradient across the filament axis, which can be interpreted
as rotation around the axis.  The rotation velocity is comparable to
the velocity dispersion.  This implies that the cloud is supported
against gravity in part by rotation, and that the centrifugal force
is as strong as the turbulent pressure, $ ( \beta \, \approx \, 1 ) $.

\bigskip

\bigskip

     The authors thank Professors Takenori Nakano and Satoshi Yamamoto
for discussions and comments.  This work is financially supported in
part by the Grant-in-Aid for General Scientific Research (05640308).

\section*{Appendix 1. \ \ The Matrix Elements of {\boldmath A}}

     Matrix $ \mbox{\boldmath $A$} $ is expressed as
\begin{equation}
{d \over d r }\mbox{\boldmath $y$} \; = \; \mbox{\boldmath $A y$},
\end{equation}
\begin{equation}
(y_1, \, y_2, \, y_3, \, y_4) \; = \;
\left( P_1 \; + \; { \mbox{\boldmath $B$}_0 \cdot
\mbox{\boldmath $B$}_1 \over 4 \pi } , \,
{ i \rho_0 v_{1r} \over \omega }, \, \rho_{\rm c} \psi_0 , \, \rho_{\rm c} g_0
\right),
\end{equation}
\begin{equation}
\mbox{\boldmath $A$} \;
= \; \mbox{\boldmath $R$} \; + \; \mbox{\boldmath $S$	}
\mbox{\boldmath $T$}^{-1}\mbox{\boldmath $U$ },
\end{equation}
\begin{equation}
\mbox{\boldmath $R$} \; = \;
\left( \matrix{ R_{11}&R_{12}&R_{13}&R_{14} \cr
                                R_{21}&R_{22}&R_{23}&0      \cr
                                0     &0     &0     &1      \cr
                                R_{41}&0     &R_{43}&R_{44} \cr}
                \right),
\end{equation}
\begin{eqnarray}
R_{11} \; & = & \; { 1 \over {c_{\rm s}}^2 }
\left( {{ v_{ 0 \varphi }}^2 \over r } \; -  \; g_0 \right)
\; + \; { 2 v_{0 \varphi }  \over r \xi } { m \over r }, \\
R_{12} \; & = & \; \omega \xi
\; - \;
{ ( \mbox{\boldmath $k$} \cdot \mbox{\boldmath $B$}_0 )^2 \over 4 \pi \rho_0 }
{ \omega \over \xi }
\\ & \; \; \; \; &
- \,
{ 2 v_{0 \varphi } \over r }{\omega \over \xi }
\left[
{1 \over r } {d \over d r } ( r v_{0 \varphi })
\; + \;
{\mbox{\boldmath $k$ } \cdot
\mbox{\boldmath $B$}_0 \over 4 \pi \rho_0 } { 1 \over \xi }
{1 \over r } {d \over d r } ( r B_{0 \varphi })
\right] ,  \\
R_{13} \; &=& \; {2 v_{0 \varphi }  \over r \xi } { m \over r }
{ \rho_0 \over \rho_{\rm c} }, \\
R_{14} \; &=& \; - { \rho_0 \over \rho_{\rm c} }, \\
R_{21} \; &= & \; - {1 \over {c_{\rm s}}^2}{\xi \over \omega }
\; + \;
{| \mbox{\boldmath $k$}|^2 \over \omega \xi }, \\
R_{22} \; & = & \; -{1 \over r}
\; - \;
{ \mbox{\boldmath $k$} \cdot \mbox{\boldmath $B$}_0
\over 4 \pi \rho_0 } { 1 \over \xi^2 }
\left[
{m \over r }
{1 \over r } {d \over d r } ( r B_{0 \varphi })
\; + \;
k_z
{ d B_{0 z} \over d r }
\right] \nonumber
\\ & \; \; \;  &
- \,
{m \over r }{1 \over \xi }
{1 \over r } {d \over d r } ( r v_{0 \varphi }), \\
R_{23} \; & = & \;
{| \mbox{\boldmath $k$} |^2 \over \omega \xi }
{ \rho_0 \over \rho_{\rm c} },  \\
R_{41} \; &= & \; { 4 \pi G \rho_{\rm c} \over { c_{\rm s} }^2 }, \\
R_{43} \; &= & \; \vert\mbox{ \boldmath $k$} \vert ^2, \\
R_{44} \; &= & \; - \, { 1 \over r},
\end{eqnarray}

\begin{equation}
\mbox{\boldmath $S$} \; = \; \left( \matrix{ S_{11}&S_{12}\cr
                                S_{21}&S_{22}\cr
                                0     &0     \cr
                                S_{41}&S_{42}\cr } \right) ,
\end{equation}
\begin{eqnarray}
S_{11} \; & = & \; - \left[ {1 \over 2 \pi r}
\; + \;
{ 1 \over 4 \pi {c_{\rm s}}^2 }
\left( { {v_{0 \varphi }}^2 \over r} \; - \; g_0 \right) \right]
B_{0 \varphi }
\; - \;
{2 v_{0 \varphi } \over r \xi }
{ \mbox{\boldmath $k$} \cdot \mbox{\boldmath $B$} \over 4 \pi }, \\
S_{12} \; & = & \; -{ 1 \over 4 \pi {c_{\rm s}}^2 }
\left( {{v_{0 \varphi }}^2 \over r } \; -  \; g_0 \right)
B_{0 z}, \\
S_{21} \; & = & \; {1 \over  4 \pi {c_{\rm s} }^2}
{\xi \over \omega } B_{0 \varphi }
\; - \; {m \over r} {1 \over \xi \omega }
{ \mbox{\boldmath $k$} \cdot \mbox{\boldmath $B$}_0 \over 4 \pi }, \\
S_{22} \; & = & \; {1 \over  4 \pi {c_{\rm s} }^2}{\xi \over \omega } B_{0 z}
\; - \;
k_z {1 \over \xi \omega }
{ \mbox{\boldmath $k$} \cdot \mbox{ \boldmath $B$}_0 \over 4 \pi }, \\
S_{41} \; &= & \; -{G \over {c_{\rm s}}^2 } B_{0 \varphi },
\\
S_{42} \; &= & \; -{G \over {c_{\rm s}}^2 } B_{0 z},
\end{eqnarray}

\begin{equation}
\mbox{\boldmath $T$} \; = \; \left( \matrix{ T_{11} & T_{12} \cr
                                T_{21} & T_{22} \cr } \right),
\end{equation}

\begin{eqnarray}
T_{11} \; &= & \; T_{22}
\; = \; { B_{0 \varphi } B_{0 z} \over 4 \pi \rho_0 {c_{\rm s} }^2}, \\
T_{12} \; &= &\; 1 \; + \; { {B_{0z}}^2 \over 4 \pi \rho_0 { c_{\rm s} }^2 }
\; - \;
{ ( \mbox{ \boldmath $k$} \cdot \mbox{\boldmath $B$}_0 )^2
\over 4 \pi \rho_0 \xi^2 }, \\
T_{21} \; & = & \; 1 \; + \;
{ {B_{0 \varphi }}^2 \over 4 \pi \rho_0 { c_{\rm s} }^2 }
\; - \; { ( { \bf k }
\cdot \mbox{\boldmath $B$}_0 )^2 \over 4 \pi \rho_0 \xi^2 },
\end{eqnarray}

\begin{equation}
\mbox{\boldmath $U$} \; = \; \left( \matrix{ U_{11}&U_{12}&U_{13}&0\cr
                                 U_{21}&U_{22}&U_{23}&0\cr } \right),
\end{equation}
\begin{eqnarray}
U_{11} \; & =
& \; { B_{0z} \over {c_{\rm s}}^2 \rho_0 }
\; - \; k_z { \mbox{\boldmath $k$} \cdot
\mbox{\boldmath $B$}_0 \over \rho_0 \xi^2 }, \\
U_{12} \; & = & \; \left( {B_{0z} \over \rho_0 }
{ d \rho_0 \over d r}
\; - \; {d B_{0z} \over d r} \right)
{\omega \over \rho_0 \xi }
\; + \;
{ \omega (\mbox{\boldmath $k$} \cdot
\mbox{\boldmath $B$}_0 )^2 \over 4 \pi {\rho_0}^2 \xi^3}
{d B_{0z} \over d r}, \\
U_{13} \; & = & \; -k_z
{ \mbox{\boldmath $k$} \cdot \mbox{\boldmath $B$}_0 \over \xi^2 \rho_{\rm c} },
\\
U_{21} \; & = & \; { B_{0 \varphi } \over {c_{\rm s}}^2 \rho_0 }
\; - \; { m \over r }
{ \mbox{\boldmath $k$} \cdot \mbox{\boldmath $B$}_0 \over \rho_0 \xi^2 }, \\
U_{22} \; & = & \; \left( {B_{0 \varphi } \over \rho_0 }
{ d \rho_0 \over d r}
\; - \; {d B_{0 \varphi } \over d r}
\; + \; {B_{0 \varphi } \over r } \right)
{\omega \over \rho_0 \xi }
\; + \;
{ \omega (\mbox{\boldmath $k$}
\cdot \mbox{\boldmath $B$}_0)^2 \over 4 \pi {\rho_0}^2 \xi^3}
{1 \over r } {d \over d r } ( r B_{0 \varphi }) \\
 & \; \; \; &
\; + \; {\omega \over \rho_0 }
\mbox{\boldmath $k$} \cdot \mbox{\boldmath $B$}_0
\left[ {1 \over \xi^2 }
\left( {d v_{0 \varphi } \over d r} \; + \;
{v_{0 \varphi } \over r} \right)
\; - \;
{1 \over \omega^2 }
\left( {d v_{0 \varphi } \over d r} \; - \;
{v_{0 \varphi } \over r} \right) \right], \\
U_{23} \; & = & \; -{ m \over r }
{ \mbox{\boldmath $k$} \cdot \mbox{\boldmath $B$}_0 \over \rho_{\rm c} \xi^2},
\end{eqnarray}
\begin{eqnarray}
\xi \; & = & \; \omega \; - \; {m \over r} v_{0 \varphi }, \\
{\bf k} \; & = & \; \left( 0, \, { m \over r }, \, k_z \right), \\
{\bf B}_0 \; & = & \; (0,\, B_{0 ,\varphi }, \, B_{0 z} ), \\
g_0 \; &= \; d \psi_0 / dr ,
\end{eqnarray}
\begin{equation}
B _{1r} \; = \; { i \omega \over \xi \rho _0 } \,
\mbox{\boldmath $k \cdot B $}_0
\, y _2 \; ,
\end{equation}
\begin{equation}
\left( \matrix{ B_{1 \varphi } \cr B_{1 z} \cr } \right)
\; = \; \mbox{\boldmath $T$}^{-1} \mbox{\boldmath $U$}
\left( \matrix{ y_1 \cr y_2 \cr y_3 \cr y_4 \cr } \right),
\end{equation}
\begin{equation}
v _{1r} \, = \, { \omega \over i \rho _0 } \, y _1 \; ,
\end{equation}
\begin{equation}
\left( \matrix{ v_{1 \varphi } \cr v_{1 z} \cr } \right)
\; = \; \left[ \mbox{\boldmath $W$ } \; - \;
{ \mbox{\boldmath $k$} \cdot \mbox{\boldmath $B$}_0 \over 4 \pi \rho_0 \xi }
\mbox{\boldmath $T$}^{-1} \mbox{\boldmath $U$ } \right]
\left( \matrix{ y_1 \cr y_2 \cr y_3 \cr y_4 \cr } \right),
\end{equation}
\begin{equation}
\mbox{\boldmath $W$} \; = \; \left( \matrix{W_{11}&W_{12}&W_{13}&0\cr
                                W_{21}&W_{22}&W_{23}&0\cr } \right),
\end{equation}
\begin{eqnarray}
W_{11} \; & = & \; {m \over r}{1 \over \rho_0 \xi }, \\
W_{12} \; & = & \; -{\omega \over \xi \rho_0 r }\left[
{d \over d r } ( r v_{0 \varphi })
\; + \; {\mbox{\boldmath $k$} \cdot \mbox{\boldmath $B$}_0
\over 4 \pi \rho_0 \xi }
{d \over d r } ( r B_{0 \varphi })\right], \\
W_{13} \; & = & \; {m \over r}{ 1 \over \xi \rho_{\rm c} }, \\
W_{21} \; & = & \; k_z {1 \over \rho_0 \xi }, \\
W_{22} \; & = & \; - \, { \omega \mbox{\boldmath $k$} \cdot
\mbox{\boldmath $B$}_0 \over
4 \pi \rho_0{}^2 \xi^2 } {d  B_{0z} \over d r } \; ,
\end{eqnarray}
and
\begin{equation}
W_{23} = \; k_z { 1 \over \xi \rho_{\rm c} }.
\end{equation}

\noindent {\bf Appendix 2. \ \ The Bisection Method for Complex Eigenvalues}

       In this appendix we describe the method used to search for a complex
eigen
value.  According to Paper I, the condition for an eigenvalue is
expressed as
\begin{equation}
\chi ( \omega ) \, \equiv \, {\rm det} \,
\pmatrix{ y _1 ^{(1)} ( r ; \, \omega ) & y _1 ^{(2)} ( r ; \, \omega ) &
y _1 ^{(3)} ( r ; \, \omega ) & y _1 ^{(4)} ( r ; \, \omega ) & \cr
y _2 ^{(1)} ( r ; \, \omega ) & y _2 ^{(2)} ( r ; \, \omega ) &
y _2 ^{(3)} ( r ; \, \omega ) & y _2 ^{(4)} ( r ; \, \omega ) & \cr
y _3 ^{(1)} ( r ; \, \omega ) & y _3 ^{(2)} ( r ; \, \omega ) &
y _3 ^{(3)} ( r ; \, \omega ) & y _3 ^{(4)} ( r ; \, \omega ) & \cr
y _4 ^{(1)} ( r ; \, \omega ) & y _4 ^{(2)} ( r ; \, \omega ) &
y _4 ^{(3)} ( r ; \, \omega ) & y _4 ^{(4)} ( r ; \, \omega ) & \cr }
\, = \, 0 \; ,
\end{equation}
where $ \lbrack
y _1 ^{(i)} , \, y _2 ^{(i)} , \, y _3 ^{(i)} , \, y _4 ^{(i)}
\rbrack $
for $ i \, = \, 1 $, 2, 3, and 4 denote independent solutions satisfying
either of the boundary conditions at $ r \, = \, 0 $ or $ \infty $.
We first evaluate $ \chi ( \omega ) $ for $ \omega \, = \, \omega _1 $,
$ \omega _2 $, and $ \omega _3 $.  The points $ \omega \, = \,
\omega _1 $, $ \omega _2 $, and $ \omega _3 $ are on the vertexes of
a right isosceles triangle on the complex $ \omega $ plane [ see figure (13)].
We next compute
\begin{equation}
\eta ( \omega _1 \, \, \omega _2 \, \, \omega _3 ) \, \equiv \,
{\rm log} \lbrack \chi ( \omega _1 ) / \chi ( \omega _2 ) \rbrack \, + \,
{\rm log} \lbrack \chi ( \omega _2 ) / \chi ( \omega _3 ) \rbrack \, + \,
{\rm log} \lbrack \chi ( \omega _3 ) / \chi ( \omega _1 ) \rbrack \; ,
\end{equation}
where each logarithm operation takes the principal value.
The value of $ \eta ( \omega _1 \, \, \omega _2 \, \, \omega _3 ) $
can take either 0 or $ \pm 2 \pi i $.  When $ \eta \, = \, 2 \pi i $,
there is a zero point of $ \chi ( \omega ) $ inside the triangle.
Note that $ \eta \, = \, \oint \, (1/\chi) \, d \chi $.

     When $ \eta \, = \, 2 \pi i $ for the triangle $ \omega _1 -
\omega _2 - \omega _3 $, we integrate the perturbation equation for
$ \omega _4 \, = \, ( \omega _2 \, + \, \omega _3 ) / 2 $ and compute
$ \chi ( \omega _4 ) $.  Either $ \eta ( \omega _1 , \,
\omega _2 , \, \omega _4 ) $, $ \eta ( \omega _1 , \, \omega _4 , \,
\omega _3 ) $, or $ \eta ( \omega _3, \, \omega _4 , \, \omega _2 ) $
is identical to $ \eta ( \omega _1 , \, \omega _2 , \, \omega _3 ) $,
\begin{equation}
\eta ( \omega _1 , \, \omega _2 , \, \omega _3 ) \, = \,
\eta ( \omega _1 , \, \omega _2 , \, \omega _4 ) \, + \,
\eta ( \omega _1 , \, \omega _4 , \, \omega _3 ) \, + \,
\eta ( \omega _3 , \, \omega _4 , \, \omega _2 ) \; .
\end{equation}
In most cases, we obtain $ \eta ( \omega _3 , \, \omega _4 , \, \omega _2 ) \,
= \, 0 $.
[ The frequency of $ \eta ( \omega _3 , \, \omega _4 , \, \omega _2 )
\, \not= \, 0 $ is very small when the triangle
$ \omega _1 - \omega _2 - \omega _3 $ is small ].
Then, the
region containing the zero point of $ \chi \, ( \omega ) \, = \, 0 $
is reduced to half in area.  The reduced region is again
a right isosceles triangle and can be squeezed by successive
reduction.  Finally, we can obtain an eigenvalue, $ \omega $, with a
sufficiently small error.  Each iteration makes the estimated
error smaller by a factor of $ \sqrt{2} $ in this extended
bisection method.

\clearpage

\begin{figure}
\caption{Dispersion relation for models with
$ ( \alpha , \, \beta ) $  =  $ ( 0 , \, 1) $.  The thick curve
denotes the growth rate, $ \omega _i $, as a function of the
wavenumber, $ k _z H $.  The thin and dashed curves denote the
growth rate of the most unstable mode with $ ( \alpha , \, \beta , \,
\theta ) $ = $ ( 1 , \, 0 , \, 0 ^\circ ) $
and $ ( 0 , \, 0 , \, 0 ^\circ ) $, respectively, for comparison.}
\end{figure}

\begin{figure}
\caption{Dependence of the growth rate, $ \omega _i $, of
the sausage $ ( m \, = \, 0 ) $ mode
on $ \beta $ for $ \alpha \, = \, 0 $.  Each curve denotes the
growth rate of the unstable sausage mode with $ ( \alpha , \, \beta ) $
= $ ( 0 , \, 0 ) $, $ ( 0 , \, 1 ) $, and $ ( 0 , \, 2 ) $.
The ordinate is the growth rate, $ \omega _i $, and the wave number,
$ k _z $.}
\end{figure}

\begin{figure}
\caption{Wave number of the most unstable perturbation
as a function of $ \alpha \, + \, \beta $
for $ \theta \, = \, 0 ^\circ $.  The ordinate and abscissa of
the upper panel are  the wavenumbers of the most unstable perturbation,
$ k _{\rm max} H $ and $ \alpha \, + \, \beta $, respectively.
The filled circles denote the numerically obtained data points, and
the curve denotes the fitting formula, equation (22).  The lower panel
shows the error of the fitting formula.  The ordinate is
(error) $ \equiv \, $ (fitting formula)/(numerical data) $-$ 1.}
\end{figure}

\begin{figure}
\caption{Cross section of the model
filamentary cloud perturbed
by the fastest growing sausage mode (main panel).  The model parameters of the
equilibrium model are $ ( \alpha , \, \beta , \, \theta ) $ =
$ ( 0, \, 1, \, 0 ^\circ ) $. The abscissa is the $ z $-axis and the
ordinate is the radial direction.  The density is indicated by greyness,
the scale of which is shown on the left side of the panel.  The arrows denote
the velocity field on the $ r \, - \, z $ plane.  The right-hand panel shows
$ v _{\varphi} $ as a function of $ r $ at $ z \, = \, 0.0 $.
The dashed and solid curves denote the values in equilibrium and in
the perturbed state, respectively.  The growth rate and the wave
number of this perturbation are $ \omega _i \, = \,
0.606 $ and $ k _z \, = \, 0.467 $, respectively.}
\end{figure}

\begin{figure}
\caption{Same as figure 4, but for $ ( \alpha , \, \beta , \,
\theta ) $ = $ ( 0.5 , \, 0.5, \, 0 ^\circ ) $.
The growth rate and the wave number of this perturbation are
$ \omega _i \, = \, 0.640 $ and $ k _z \, = \, 0.478 $, respectively.}
\end{figure}

\begin{figure}
\caption{Dispersion relation of the kink $ ( m \, = \, 1 ) $ mode
for $ ( \alpha , \, \beta ) $ = $ ( 0 , \, 1 ) $.  The dashed and
solid curves denote $ \omega _r $ and $ \omega _i $,
the oscillation frequency and growth rate, respectively.
The three horizontal dotted lines denote the angular velocities, $ \Omega $,
at $ r \, = \,2, \,  2 \surd 2 $, and 4,respectively.}
\end{figure}

\begin{figure}
\caption{Same as figure 6, except for $ ( \alpha , \, \beta ) $ =
$ ( 0, \, 0.5 ) $.}
\end{figure}

\begin{figure}
\caption{Same as figure 6, except for $ ( \alpha , \, \beta ) $ =
$ ( 0, \, 2 ) $. }
\end{figure}

\begin{figure}
\caption{Same as figure 4, except for the most unstable kink
mode with $ ( \alpha , \, \beta ) \, = \, ( 0 , \, 1 ) $.
The wave number and growth rate are $ H k {}_z \, = \, 0.323 $
and $ \omega _{i, \, {\rm max}} \, = \, 0.135 \, \surd(2 \pi G \rho _{\rm c})$,
respectively.  The amplitude of the perturbation is normalized
so that $ \rho _1 / \rho _{\rm c} \, = \, 0.5 $ at maximum.
(a) The cross section in the $ r \, - \, z $ plane. (b) The cross section
 of $ z \, = \, 0 $ .}
\end{figure}

\begin{figure}
\caption{Same as figure 6, except for the kink mode of
$ ( \alpha , \, \beta , \, \theta ) \,
= \, ( 0.5, \, 0.5, \, 0 ^\circ ) $.}
\end{figure}

\begin{figure}
\caption{Dispersion relation of the sausage mode with $ ( \alpha ,
\, \beta ) \, = \, ( 0.5, \, 0.5 ) $.  The curves denote the
growth rate as a function of $ k _z $ for $ \theta \, = \, 30 ^\circ $,
$ 60 ^\circ $, and $ 90 ^\circ $.  Figures 11 (a) and (b) show
$ \omega _r $ and $ \omega _i $, respectively.
The dotted curve denotes $ \omega _i $ for a model with
$ \theta \, = \, 0 ^\circ $ for comparison.}
\end{figure}

\begin{figure}
\caption{Same as figure 4, except for the sausage mode with
$ ( \alpha , \, \beta , \, \theta ) \, = \, ( 0.5, \, 0.5, \, 60 ^\circ ) $.
The growth rate and wave number are $ \omega \, = \,
(0.0199 \, + \, 0.594 \, i )
\, \surd(2 \pi G \rho _{\rm c}) $ and $ k _z H \, = \, 0.432 $,
respectively.}
\end{figure}

\begin{figure}
\caption{Search for a complex eigenvalue on the complex $ \omega $ plane.
The region for the search narrows by a factor of 2 in area by each iteration.}
\end{figure}


\bigskip


\clearpage

\begin{references}

\reference
Balbus S.~A.,  Hawley, J.~H. 1991, {ApJ} {376}, 214

\reference
Bally J. 1989, in {Low Mass Star Fromation and Premain Sequence
Objects}, ed Reipurth B. (ESO, Garching) p1


\reference
Bonnell I., Arcoragi J.-P., Martel H., Bastien P.
1992, {ApJ} {400}, 579



\reference
Chandrasekhar S.,  Fermi E. 1953, {ApJ} {118}, 116



\reference
Fujimoto M.~Y. 1987, {A\&A} {176}, 53.


\reference
Glatzel W. 1987a, {MNRAS} {225}, 227

\reference
Glatzel W. 1987b, {MNRAS} {228}, 77

\reference
Habe A., Uchida  Y., Ikeuchi  S., Pudritz  R.E. 1991,
{PASJ}  {43}, 703

\reference
Hanawa  T. 1986,  {MNRAS}  {223}, 859

\reference
Hanawa  T. 1987, {A\&A}  {179}, 383

\reference
Hanawa  T., Nakamura  F., Matsumoto  T., Nakano  T., Tatematsu  K.,
Umemoto  T., Kameya  O. et al. 1993, {ApJL}  {404}, L83

\reference
Hansen  C.~J., Aizenman M.~L.,  Ross R.~R. 1976,
{ApJ} {207}, 736


\reference
Inutsuka S.,  Miyama S. M. 1992, {ApJ} {388}, 392



\reference
Kato S. 1987,  {PASJ} {39}, 645


\reference
Lin C. C. 1945a, {Quart. Appl. Math.} {3}, 117

\reference
Lin C. C. 1945b, {Quart. Appl. Math.} {3}, 218




\reference
Moneti A., Pipher J.~L., Helfer H.~L., McMillan R.~S., Perry M.~L. 1984,
{ApJ} {282}, 508



\reference
Nagasawa M. 1987, {Prog. Theor. Phys.} {77}, 635



\reference
Nakamura F., Hanawa T.,  Nakano T. 1993,  {PASJ}
{45}, 551 (Paper I)


\reference
Olano C.~A., Walmsley C.~M.,  Wilson T.~L. 1988,
{A\&A} {196}, 194

\reference
Ostriker J. 1964, {ApJ} {140}, 1056



\reference
Stodo\l kiewicz J. S. 1963, {Acta Astron.} {13}, 30

\reference
Sung C.~H. 1974, {A\&A} {33}, 99.



\reference
Tatematsu K., Umemoto T., Kameya O., Hirano N., Hasegawa T.,
Hayashi M., Iwata T., Kaifu N. et al.
1993, {ApJ} {404}, 643

\reference
Tomisaka K. 1993, {ApJ} submitted to

\reference
Uchida Y., Fukui Y., Minoshima Y., Mizuno A., Iwata T., Takaba H.
1991, {Nature} {349}, 140

\reference
Vrba F.~J., Strom S.~E., Strom K.~M. 1976,
{AJ} {81}, 958


\end{references}
\end{document}